\pdfoutput=1

\documentclass[preprint,authoryear,12pt]{elsarticle}

\usepackage{graphicx}
\usepackage{amssymb}

\journal{Advances in Space Research}

\begin{document}


\begin{frontmatter}

\title{Coronal partings}

\author[GAISh]{Igor F. Nikulin\corref{cor}}
\cortext[cor]{Corresponding author}
\ead{ifn@sai.msu.ru}

\author[GAISh,IKI,MPIPKS]{Yurii V. Dumin}
\ead{dumin@yahoo.com}

\address[GAISh]{P.K.~Sternberg Astronomical Institute
                of M.V.~Lomonosov Moscow State University,\\
                Universitetskii prosp.\ 13, 119234, Moscow, Russia}

\address[IKI]{Space Research Institute
              of Russian Academy of Sciences,\\
              Profsoyuznaya str.\ 84/32, 117997, Moscow, Russia}

\address[MPIPKS]{Max Planck Institute for the Physics of Complex Systems,\\
                 Noethnitzer Str.\ 38, 01187, Dresden, Germany}

\begin{abstract}

The basic observational properties of ``coronal partings''---the special
type of quasi-one-dimensional magnetic structures, identified by a comparison
of the coronal X-ray and EUV images with solar magnetograms---are
investigated.
They represent the channels of opposite polarity inside the unipolar
large-scale magnetic fields, formed by the rows of magnetic arcs directed
to the neighboring sources of the background polarity.
The most important characteristics of the partings are discussed.
It can be naturally assumed that---from the evolutionary and spatial
points of view---the partings can transform into the coronal holes and
\textit{visa versa}.
The classes of global, intersecting, and complex partings are identified.

\end{abstract}


\end{frontmatter}

\parindent=0.5 cm


\section{Introduction}
\label{sec:Intro}

X-ray investigations of the Sun in the recent decades revealed
a number of specific coronal structures, such as the bright X-ray points
\citep{Golub74} and coronal holes
\citep{Altschuler72,Fuerst75,Timothy75,Nolte76}.
The coronal holes (CH) attracted especial attention after discovery of
their geoefficiency \citep{Krieger73,Neupert74,Nolte76}.
However, genetic and spatial relations between CH and other
coronal structures and magnetic fields are still investigated poorly.
So, it is interesting to look for the additional large-scale
morphological features in the solar corona and to reveal their
relation to the previously-known ones.

With this aim in view, during the last 15~years we performed a visual
inspection of a large number of images regularly provided by
the leading space- and ground-based observatories.
First of all, we used the data by
\textit{Yohkoh} SXT\footnote{
The most part of original data utilized in the present paper were
downloaded from the archive at {\tt http://www.lmsal.com/SXT},
which was later relocated to another address:
{\tt http://ylstone.physics.montana.edu/ylegacy/}.}
\citep{Tsuneta91} and
\textit{SOHO} EIT \citep{Delaboudiniere95}.
In the recent time, we began to employ also the very-high-quality
images by \textit{SDO} AIA \citep{Pesnell12}.
To confront the coronal structures visible in soft X-rays and
extreme ultraviolet (EUV) with the associated magnetic fields,
we used the observations of photospheric magnetic fields mostly
by \textit{SOHO} MDI\footnote{
The original data are available in the archive:
{\tt http://sohowww.nascom.nasa.gov}.}
\citep{Scherrer95}
and, recently, by \textit{SDO} HMI \citep{Pesnell12}.
Synoptic magnetic maps by the ground-based observatories were
also occasionally employed.
In total, we have analyzed over 4000 spectroheliograms and magnetograms,
starting from 1999.\footnote{
Unfortunately, the so long period of data collection resulted in
some nonuniformity of the graphic formats as well as a number of
parasitic designations in the images presented below.}

\begin{figure}
\begin{center}
\includegraphics[width=13.7cm]{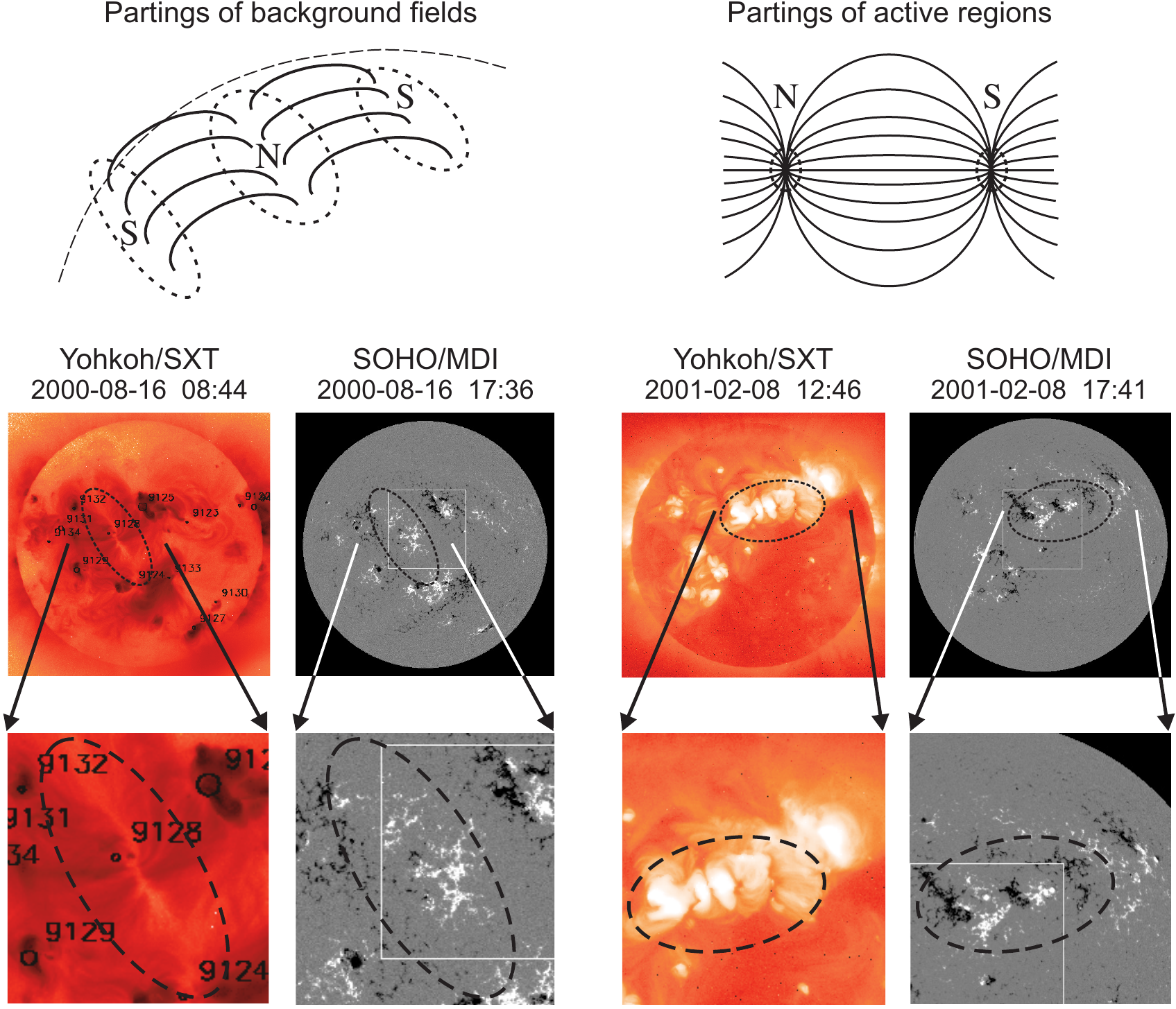}
\end{center}
\caption{Scheme of the partings of background fields
with an example for 16 August 2000 (left-hand side)
and of the active regions with an example for
08 February 2001 (right-hand side).
The patterns of soft X-ray emission (left column) and
photospheric magnetic field (right column) are given
for each case.
The images by \textit{Yohkoh}\/ SXT and \textit{SOHO} MDI
involve a number of irrelevant designations, which should not
be taken into account.
Besides, \textit{Yohkoh}\/ SXT images in the left- and
right-hand sides are in the inversed palettes (negative and
positive).
\label{figure1}}
\end{figure}

In particular, we searched for the characteristic elongated X-ray and
ultraviolet structures with a reduced brightness, which were
associated with the unipolar regions.
(It is necessary to discriminate them from the filaments,
which are localized at the boundaries between different polarities.)
As a result, it was found that the magnetic field lines, represented by
the X-ray loops, over the regions of magnetic field of the same
sign often form the specific channels between two rows of the arcs
directed oppositely, i.e.\ toward the regions of another polarity
(Fig.~\ref{figure1}).
Because of their similarity to a hair parting---where hairs play
the role of the field lines---\citet{Nikulin03} suggested to call them
the coronal partings (CP).

A few years later, \citet{Molodenskii07} justified theoretically
formation of such partings and performed their computer simulation.
They also concluded that these partings are the inevitable and natural
component of the magnetized solar atmosphere, representing the
``sign-reversal lines for the normal component of the curvature vector
of the magnetic lines''.

So, it is the aim of the present paper to give a short review
of the most important morphological features of CP, revealed in
the course of our long-term investigations, as well as
to discuss their relation to other coronal structures.

\section{Basic morphological features}

\subsection{Definitions}

The coronal partings can be subdivided into two groups:
CP of the large-scale background fields (Fig.~\ref{figure1}, left-hand side)
and the partings of active regions (AR), which represent the short and
narrow channels inside the strong fields of AR (Fig.~\ref{figure1},
right-hand side).
In such a case, the most fraction of the magnetic flux of AR and the parting
is enclosed between the leading and trailing parts of the sunspot group.
However, if there are other AR or floccules nearby, then the peripheral
part of the flux is connected to them.
As a result, a short narrow channel is formed between the deviating field
lines.
It is usually located perpendicularly to the axis of the group and
passes through the sunspots themselves.
When the sunspots disappear and the respective magnetic fields decay,
the coronal partings of active regions can be transformed into CP of
background fields.

\subsection{Distinction between the coronal partings and holes}

To emphasize that CP and CH are different structures, let us formulate
the main differences between them:
\begin{enumerate}
\item The magnetic fields of CH are weak and, at the level of
sensitivity of apparatus, are almost absent.
On the other hand, according to \textit{SOHO} MDI magnetograms,
the magnetic field of a typical CP is about 10--50\,Gs in the partings
of background fields and a few hundred Gs in the CP of active regions.
\item The coronal holes, by definition, are the coldest structures
of the corona \citep{Timothy75}, while the coronal partings possess
a higher temperature and brightness.
\item The coronal holes usually occupy large areas on the Sun and
do not have the elongated shape, characteristic of the coronal partings.
From the mathematical point of view, CP is associated with
a quasi-one-dimensional line; while CH, with a two-dimensional object
(oval).
\item The coronal holes are usually observed in the polar regions and
less frequently in the low latitudes \citep{Munro72},
while the coronal partings usually emerge in the so-called
``royal latitudes''.
\item The coronal holes are observed more frequently near the minimum of
solar activity; while the coronal partings are scarce near the minimum,
most probably, because of the weakness (low structural contrast) of
the corona and absence of the visible coronal loops, identifying CP.
\end{enumerate}

\begin{figure}
\begin{center}
\includegraphics*[width=13.7cm]{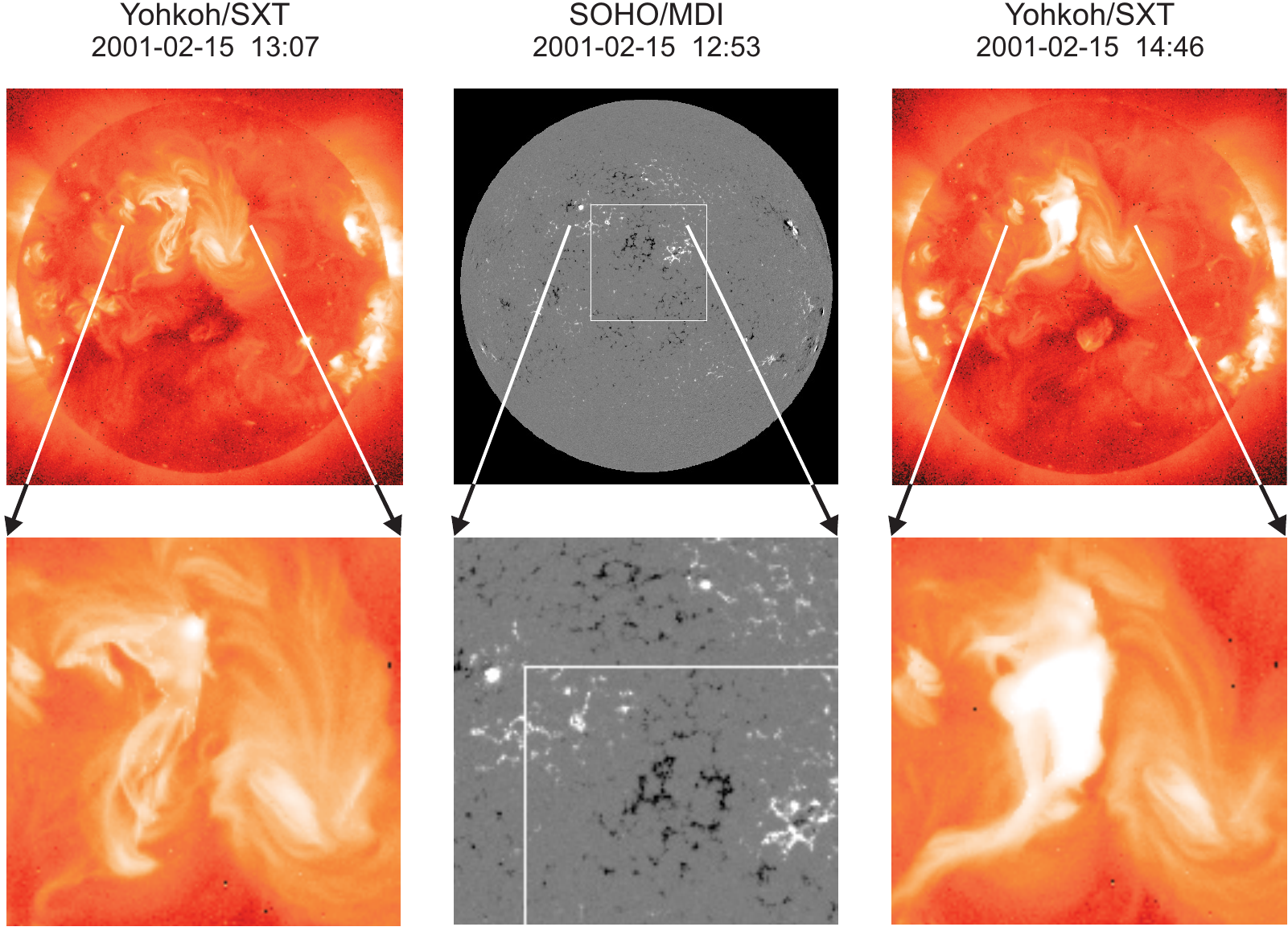}
\end{center}
\caption{Asymmetric parting in the northern hemisphere
on 15 February 2001:
the patterns of soft X-ray emission (left and right columns)
and photospheric magnetic field (central column).
\label{figure2}}
\end{figure}

Nevertheless, all these differences do not exclude the possibility of
evolution of CP into CH in the course of decay of the X-ray structures
and magnetic fields.
Moreover, the above-listed features emphasize the similarity of
these coronal structures, associated with presence of the open field
lines.

\subsection{Structure of the coronal partings}

The structure of CP and their lifetime are determined by the background
(photospheric) magnetic fields and the ambient arc formations.
Each knot in the magnetogram is a footpoint of the respective coronal arc
with lifetime about a day (and sometimes longer).
We believe that the evolution of CP is usually connected with the emergence
of small bipolar regions---the bright X-ray points (BXRP)---forming their
own system of arcs, associated with new ambient fields, and followed by
the decrease in contrast and blurring the parting.
If the CP-forming arcs weaken, then the parting expands, darkens, and
can be transformed into CH.

Roughly speaking, the X-ray brightness of CP is associated with
the magnetic-field gradients: the stronger is variation of the field,
the greater is the X-ray brightness of the corresponding structure.
Namely, the brightness is higher in the case of CP interaction
with a neighboring active region as compared to the case of the arcs
directed to a weak floccule.
As a result, the partings with substantially different brightness
at the opposite sides (i.e., the asymmetric partings) can emerge.
A typical example took place on 15 February 2001: a flare process
developed in the northern hemisphere near the central meridian,
while the asymmetric parting with an enhanced eastern part was seen
slightly to the west (Fig.~\ref{figure2}).
It can be assumed from these images that the magnetic-field arcades of
CP serve as a ``barrier'' for propagation of the flare.

As is seen in the soft X-ray images by \textit{Yohkoh},
the active-region CP (Fig.~\ref{figure1}, right-hand side) usually
passes through sunspots and separates the parts of the magnetic flux
linked to the sunspots of opposite polarity in the corresponding group
and to the peripheral magnetic fields.
The same behavior can be observed in EUV images by \textit{SDO}.

It is important that if CP intersects an equator, then sign of
the respective magnetic field does not change, as distinct from AR,
whose signs of the leading polarity are different in the opposite
hemispheres.

\subsection{Magnetic-field topology in the partings}

\begin{figure}
\begin{center}
\includegraphics*[width=13.7cm]{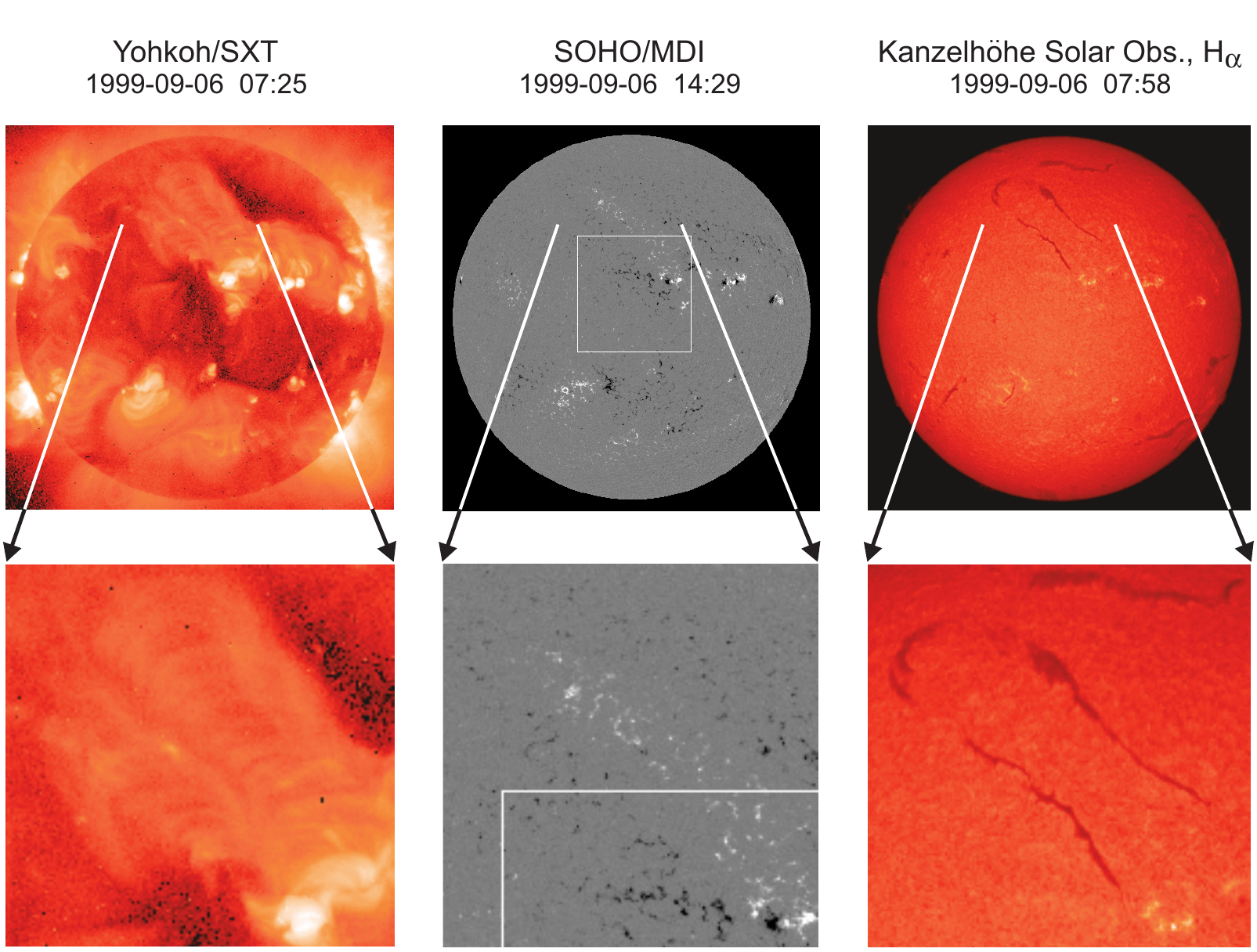}
\end{center}
\caption{Coronal parting between two filaments on
06 September 1999:
the patterns of soft X-ray emission (left column),
photospheric magnetic field (central column), and
H${}_{\alpha}$ emission (right column).
\label{figure3}}
\end{figure}

Despite the fact that CP (for example, in the images of lines of
the multiply-ionized iron, 171 and 195\,{\AA}) look like the channels
of filaments, they are absolutely different structures from
the viewpoint of magnetic topology.
The filament channels and filaments themselves are formed at the boundary
of the magnetic polarity \citep{Martres66}, i.e.\ near the neutral line
of the longitudinal magnetic field; while CP are formed in the regions of
the same sign \citep{Nikulin06}.
In other words, CP emerge between the footpoints of arcs of the same
polarity; while the filaments, at the top of these arcs, near
the neutral line.
In some cases, CP can be formed between two parallel filaments
(e.g., the interesting case on 06 September 1999 in the northern
hemisphere, presented in Fig.~\ref{figure3}).

Since CP are localized at relatively low altitudes, they are often
overlapped by the high-altitude X-ray structures.
So, when from the analysis of magnetograms CP can be assumed to exist
near the limb, they may be invisible in the X-ray images.

As regards the global magnetic topology, CP can be associated with
a separatrix.
In the simplest case of a narrow straight parting, it will represent
a plane normal to the photosphere.
Generally speaking, not every coronal parting is the separatrix
(i.e., separates two globally-distinct magnetic fluxes):
this should be proved by the magnetic field calculations in each
particular case.
And \textit{vice versa}, not every separatrix is identical to
the parting \citep[e.g.,][Ch.~4]{Somov13}:
for example, if there is a sufficiently strong horizontal component
of the magnetic field along the separatrix,
then it will hardly manifest itself as CP.
Therefore, the separatrices coincide with partings only in the
particular cases; so that CP can be reasonably considered as
the specific morphological structures in the solar corona.

\subsection{The coronal partings in different spectral lines}

The structure of CP in the infrared line of helium 10\,830\,{\AA} is
weak and smeared out: like CH, they are only a few percent brighter
than the background (as an example, see bottom left panel in
Fig.~\ref{figure4}).

In the H${}_{\alpha}$ line, the partings are identified by a weak
emission, while in the magnetograms they often exhibit themselves as
the parallel rows of unipolar magnetic knots---footpoints of the arcs
forming the partings.

Apart from the X-ray images, CP are well distinguishable in EUV lines
Fe\,XV 284\,{\AA} (e.g., right panel in Fig.~\ref{figure5}, below).
On the other hand, in the lines Fe\,X and Fe\,XII they are almost
indistinguishable from the channels of filaments.

\section{Types of the coronal partings and their evolution}

\subsection{Global coronal partings}

\begin{figure}
\begin{center}
\includegraphics*[width=13.5cm]{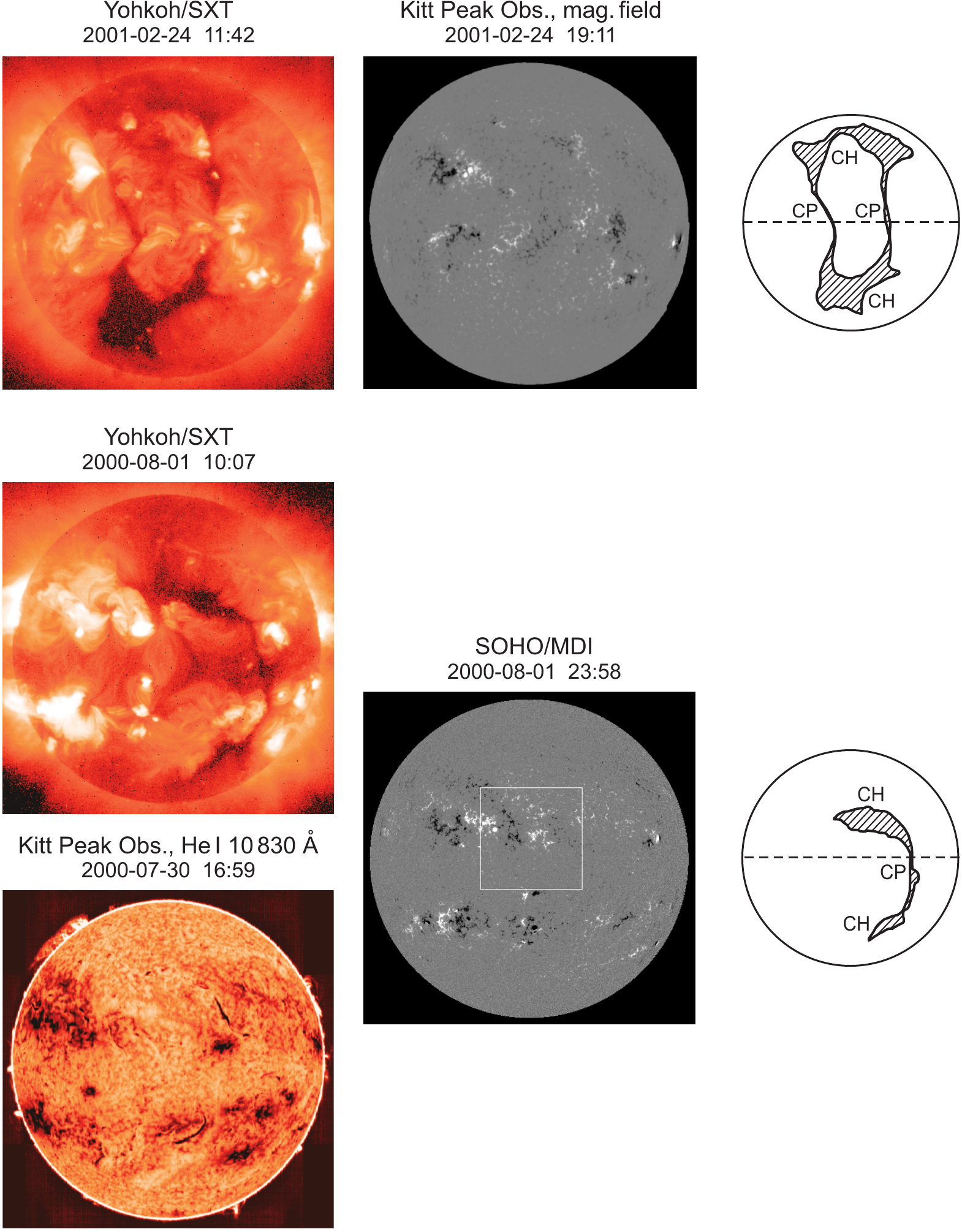}
\end{center}
\caption{The global coronal partings on
23--24 February 2001 (top) and
01 August 2000 (bottom):
the patterns of soft X-ray and infrared emission
(left column, upper and lower panels respectively),
photospheric magnetic field (central column), and
schemes of the basic coronal structures (right column).
\label{figure4}}
\end{figure}

One can often observe the global structures with a decreased X-ray brightness,
when a background or active-region CP passes from a polar coronal hole
to the CH at another pole (for example, on 23--24 February 2001 and
01 August 2000, Fig.~\ref{figure4}).
The total length of these structures is, therefore, comparable to
the solar diameter, while partings represent the links between CH
localized in the opposite hemispheres.
The lifetime of such CP is comparable to or somewhat less than
the lifetime of corresponding CH.
The low-latitude part of CP is usually placed to the west with
respect to the high-latitude part.
This seems to be caused by a differential rotation during the long
lifetime of these structures.
Two examples of the global structures in the X-ray and infrared
spectroheliograms and magnetograms are presented in Fig.~\ref{figure4}.

\subsection{Complex coronal partings}

\begin{figure}
\begin{center}
\includegraphics*[width=13.7cm]{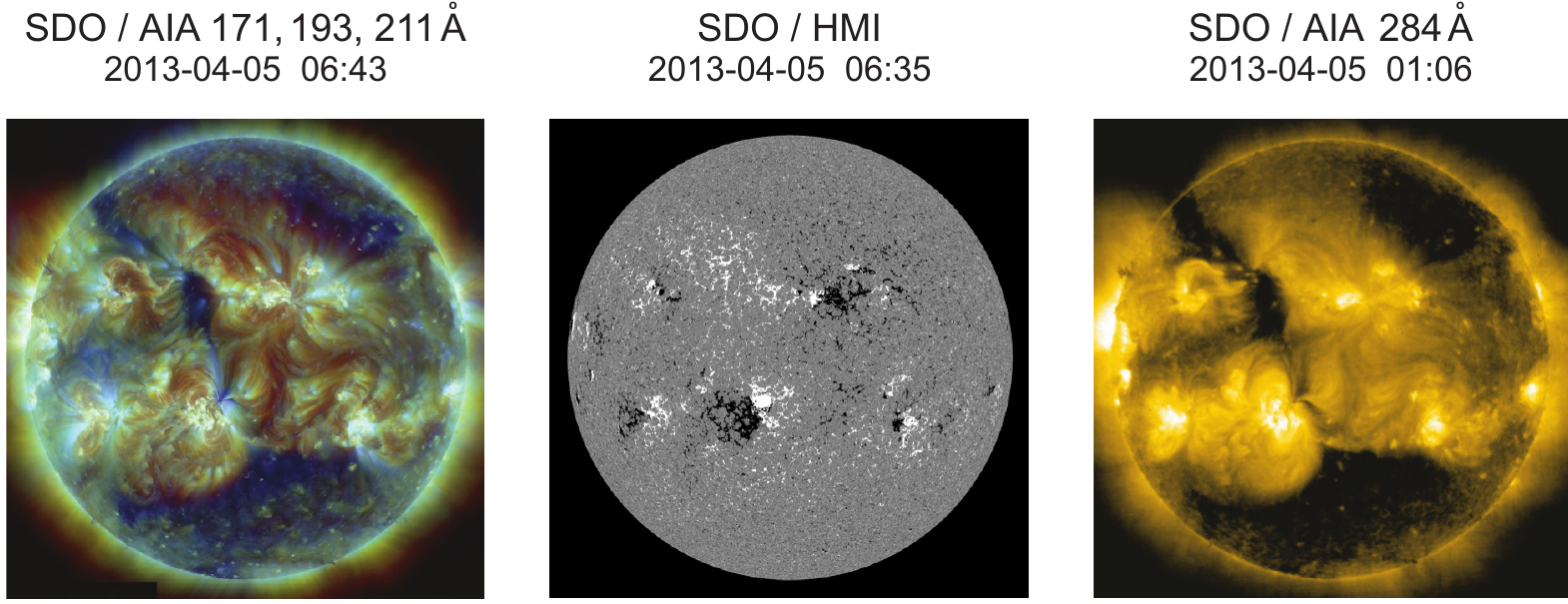}
\end{center}
\caption{The complex coronal parting on 05 April 2013:
patterns of EUV emission (left and right panels)
and photospheric magnetic field (central panel)
by \textit{SDO} satellite.
\label{figure5}}
\end{figure}

The complex (or composite) partings can be formed by both the active-region
and background CP.
As an example, we can consider the data by \textit{SDO} satellite
on 05 April 2013, presented in Fig.~\ref{figure5}:
a global parting passes from the north-east limb through the equator
and a large sunspot in the south-east and finally terminates at CH
in the southern polar region.
This parting is best seen in Fe\,XV 284\,{\AA} line.
It is composed of a few different parts:
(1)~from the NE limb to a small CH near the center of the disc at
$\sim$15E\,25N,
(2)~through this CH, after the equator, it passes between filaments
of a large sunspot,
(3)~through the sunspot and between the filaments of superpenumbra in
its southern part, up to the extensive CH in the south;
the magnetic polarity of the photospheric background remaining always
the same.

\subsection{Intersecting coronal partings}

\begin{figure}
\begin{center}
\includegraphics*[width=13.5cm]{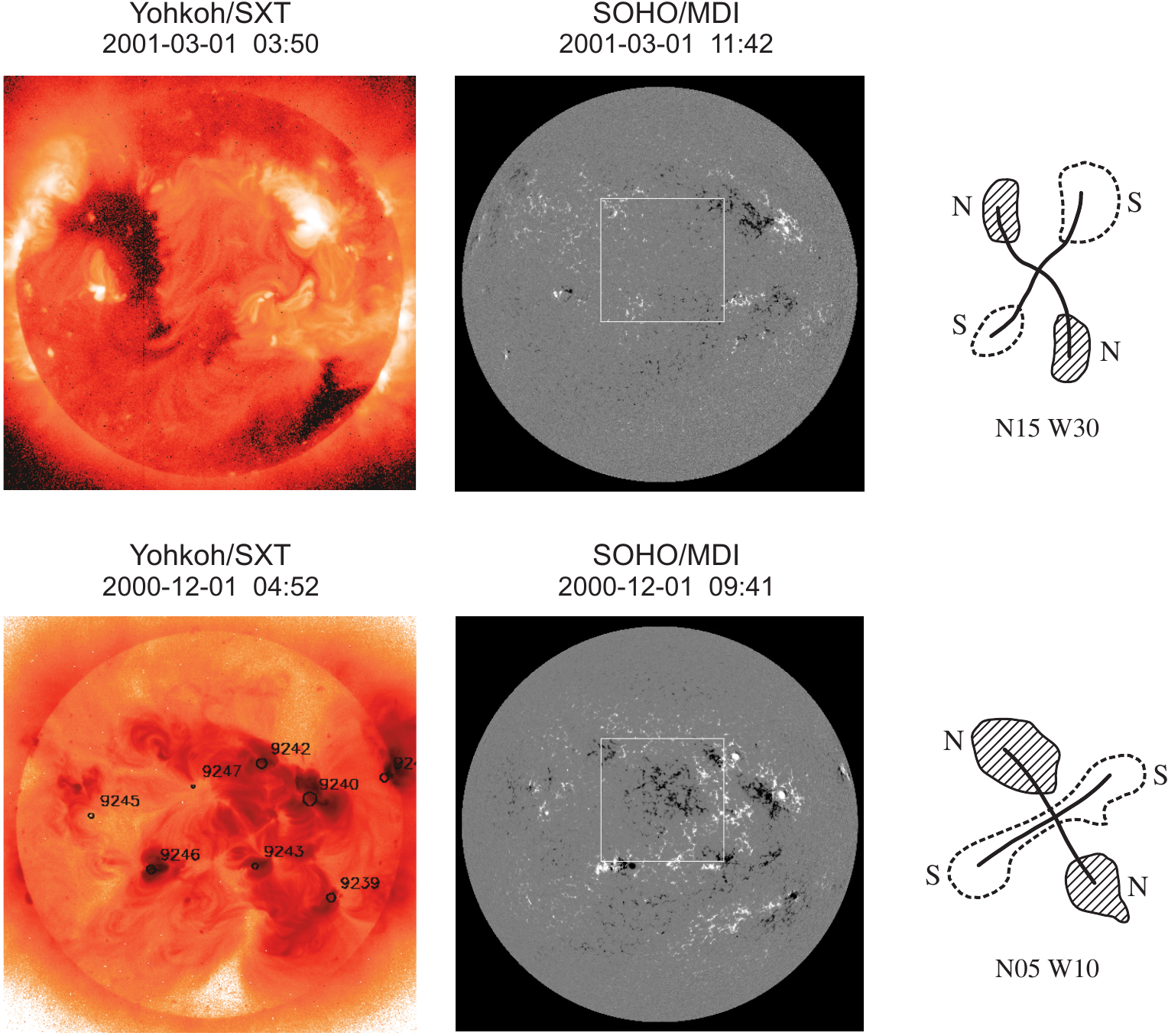}
\end{center}
\caption{The intersecting coronal partings
on 01 March 2001 at N15\,W30 (top row) and
on 01 December 2000 at N05\,W10 (bottom row):
the patterns of soft X-ray emission (left column),
photospheric magnetic field (central column), and
schemes of the basic coronal structures (right column).
The \textit{Yohkoh}\/ SXT images in the top and
bottom rows are in the inversed palettes (positive and
negative).
\label{figure6}}
\end{figure}

As was already emphasized above, each CP is associated with
a magnetic field of the same polarity.
However, the partings with different polarities can intersect each other
in the course of evolution and, as a result, the X-point should be formed
(e.g., on 01 December 2000 and 01 March 2001, as illustrated in
Fig.~\ref{figure6}).
As follows from the magnetograms, one of polarities dominates in
the point of intersection.
The role of such X-type partings in the physical processes, for example,
in the solar flares is still to be clarified.

\subsection{Evolution of the partings}

It was already mentioned earlier that CH and CP should be considered as
similar structures.
Therefore, their mutual transformation is a very important issue.
We can imagine, for example, the following evolution:
A decaying active region becomes unipolar during its long lifetime,
stretches out by the differential rotation into an approximately linear
(quasi-one-dimensional) structure, and its loops extend to
the neighboring AR of opposite polarity.
The loops originating approximately orthogonal to the surface possess
the specific thermodynamic character, because they are almost open
outwards.
As a result, they loose their density, temperature, and visibility.
On the other hand, the denser closed loops attached to the neighboring
opposite-polarity magnetic regions remain observable, i.e.\ a coronal
parting is formed.

The evolution of a large CH in the course of a few revolutions of
the Sun---from November 1972 to August 1973---was traced already in
one of the first studies of CH \citep{Timothy75}.
When the area of this hole progressively decreased, a global coronal
parting was formed in the same place by 20 June 1973 (Fig.~2b in
the above-cited article).
Next, in August 1973, the parting also began to decay, followed by
the emergence of new active regions and BXRP.
This is just the example of evolution of CH to CP, as well as
a subsequent disappearance of CP.

A detailed case study of the inverse process---transformation of
CP into CH---is still to be done.

\section{Discussion and conclusions}

A great variety of the magnetic plasma structures are realized in
the solar atmosphere.
They can experience, in fact, continuous transitions from one type
to another.

The coronal holes and partings, considered in the present paper,
possess quite similar basic features.
The most important of them is presence of the open magnetic field lines.
They determine the temperature regime and photometric characteristics
of these structures.
However, the area covered by the open field lines in CP is substantially
less than in CH.
So, their immediate influence on the formation of the solar wind should be
substantially reduced.
Nevertheless, the above-mentioned similarity between CP and CH enables
them to transform into each other, thereby affecting the space weather.

It is interesting also to mention that recent numerical simulations
by \citet{Chen15} demonstrated a surprising stability of the emissive
EUV loops as compared to the magnetic field: while the evolving field lines
successively enter and leave the narrow regions favorable for the release
of energy, the bright loops remain actually in the same place.
This is an additional argument for the importance of studying
the large-scale arcade structures in the solar corona.

\medskip

In summary, consideration of a few examples of the coronal partings
performed in the present paper suggests that they possess the following
basic properties:
\begin{enumerate}
\item A coronal parting is the elongated unipolar magnetic structure
formed by two rows of arcs, which are directed to the neighboring regions
of opposite polarity.
Since coronal partings are formed by the magnetic loops,
the structure and lifetime of CP are determined by the respective
magnetic fields.
\item There can be global partings on the Sun, which usually connect
the polar CH. Correspondingly, their characteristic size is comparable
to the solar diameter (as is seen in the particular examples presented in
Fig.~\ref{figure4}); and the period of their existence is comparable to
the CH lifetime (as follows from visual inspection of the series of
additional images, not presented here).
\item The coronal partings passing through the equator do not change
the sign of the magnetic field.
On the other hand, there can be partings of opposite polarity that
intersect each other.
\item The placement of CP and filaments are absolutely different:
the filaments are localized at the top of arcs; while CP, near their
footpoints.
\item Roughly speaking, the width and length of CP are inversely
proportional to the magnetic field intensity:
the longest CP are usually associated with weak background fields,
stretched by the differential rotation; while short and narrow CP are
associated with strong magnetic fields of the active regions
(Fig.~\ref{figure1}).
\item The system of magnetic field lines of CP covers a considerable
area and interacts with two neighboring regions of opposite polarity.
Thereby, it should play an important role in the hierarchy of the
large-scale coronal magnetic fields.
\item By now, we were unable to detect any amplification of the solar
wind associated with CP, which is not surprising: the area of open
magnetic field lines in CP is usually quite small as compared to
the coronal holes.
\end{enumerate}

\section*{Acknowledgements}

We are grateful to B.V.~Somov for valuable discussions and advises
as well as to \textit{SOHO}, \textit{Yohkoh}, and \textit{SDO}
teams for the possibility to use their data.

This work was partially supported by the Russian Foundation for
Basic Research, grant no.~08-02-01033a.

\end{document}